\documentclass[12pt,preprint]{aastex}

\slugcomment{Astrophysical Journal Letters, in press}

\shorttitle{Fermi and TeV constraints on the EBL}
\shortauthors{Georganopoulos, Finke, \& Reyes}

\begin{document}

\title{A  method for setting upper limits to  the extragalactic background light with {\em Fermi}-LAT and TeV observations 
of blazars}

\author{ Markos Georganopoulos\altaffilmark{1,2},  Justin
D. Finke\altaffilmark{3,4}, \&  Luis C. Reyes\altaffilmark{5}}

\altaffiltext{1}{Department of Physics,
University of Maryland Baltimore County, 1000 Hilltop Circle,
Baltimore, MD 21250, USA}
\altaffiltext{2}{NASA Goddard Space
Flight Center, Code 663, Greenbelt, MD 20771, USA}
\altaffiltext{3}{U.S. Naval Research Laboratory, Code 7653, 4555
Overlook Ave SW, Washington, DC 20375-5352, USA}
\altaffiltext{4}{NRL/NRC Research Associate}
\altaffiltext{5} {Kavli Institute for Cosmological Physics (KIPC) at
the University of Chicago, Chicago, IL 60637, USA}

\begin{abstract}
We propose a method for setting upper limits to the extragalactic background light (EBL). Our method uses
simultaneous {\em Fermi}-LAT  and ground-based TeV observations of
blazars and is based on the assumption that the intrinsic spectral
energy distribution (SED) of  TeV
blazars lies below the extrapolation
of the {\em Fermi}-LAT SED  from GeV to TeV energies.
 By extrapolating the {\em Fermi}-LAT spectrum, which for TeV blazars is practically
unattenuated by photon-photon pair production with EBL photons,  a
firm upper limit on
the intrinsic SED at TeV energies is provided.  
The ratio of the
extrapolated spectrum
to the observed TeV spectrum provides  upper limits
to the optical depth for the propagation of the TeV photons due to
pair production on the EBL, which in turn sets firm upper limits  to
EBL models. 
We demonstrate our method using simultaneous observations from {\em Fermi}-LAT and  ground-based TeV telescopes of the blazars \object{PKS~2155-304} and \object{1ES~1218+304}, and show that  high EBL density models are disfavored. We also discuss  how our
method can be optimized and
 how {\em Fermi } and X-ray monitoring observations of TeV blazars can
guide future TeV campaigns, leading to potentially much stronger
constraints on  EBL models.
\end{abstract}


\section{Introduction \label{section:intro}}

  
 The EBL reflects the cosmologically important time-integrated history of light production and re-processing in the Universe. For this reason, measuring its intensity is highly desirable.  
 The two
 components of the EBL are dust emission 
 peaking at $\lambda\sim 100 \mu m$ and starlight  peaking at $\sim 1 \mu m$. The actual level of the
 EBL is very difficult to measure, due to the dominance of foreground
 emission, mostly from interplanetary dust in our solar system
  \citep[for reviews see][]{hauser01,kashlinsky05}, and   
 the EBL level remains unknown within a factor of $\sim$
 few. Model-independent lower limits to the EBL, based on galaxy counts
 (e.g. Dole et al. 2006, B\'ethermin et al. 2010), are the only strict lower limits on
 the EBL to date.  Modeling the EBL can lead to definite prediction,
 however uncertainties in the star formation rate, initial mass
 function, dust extinction, and how they evolve
 with redshift, has led to significant discrepancy among models
  \citep{salamon98, stecker06, primack05, gilmore09, kneiske02,
 kneiske04, razzaque09, finke10_model, franceschini08}.
 Recently, \cite{georganopoulos08} proposed a new method based on detecting as GeV
 emission EBL radiation that has been inverse Compton-scattered by
 relativistic electrons in the lobes of nearby radio galaxies such as Fornax~A.

    
 The EBL in the 1-10 $\mu$m range can in principle be obtained
 by using the TeV blazars as background light sources and modeling its
 attenuation due to pair production with EBL photons: by assuming that
 the intrinsic TeV spectrum is known from modeling of the broadband
 blazar SED, one can derive the mid-IR EBL by comparing the observed
 to the presumed intrinsic spectrum  \citep[e.g.,][]{stecker92,stanev98,renault01}.  Because it is not possible to determine with
 confidence the intrinsic TeV spectrum, a variation
 of this method has been proposed that  sets limits on the EBL by assuming that the intrinsic
 blazar TeV spectrum cannot be arbitrarily hard:  from simple shock acceleration arguments
 one would not expect
 an intrinsic TeV photon index harder than $\Gamma_{TeV}=1.5$
 \citep[e.g.,][]{aharonian06}.  
 Detailed shock acceleration simulations, however,  indicate that
 harder VHE spectral indices may be possible \citep{stecker07}.
   A large
 lower electron Lorentz factor 
 \citep{katarzynski06}, and Compton scattering of the cosmic microwave
 background in an extended jet may also lead to hard TeV spectra  \citep{boett08}.
 These  considerations significantly relax the EBL limits derived  by  assuming $\Gamma_{TeV}\geq 1.5$  \citep{mazin07,finke09}.

\begin{figure}
\epsscale{0.5}
\includegraphics[scale=0.6]{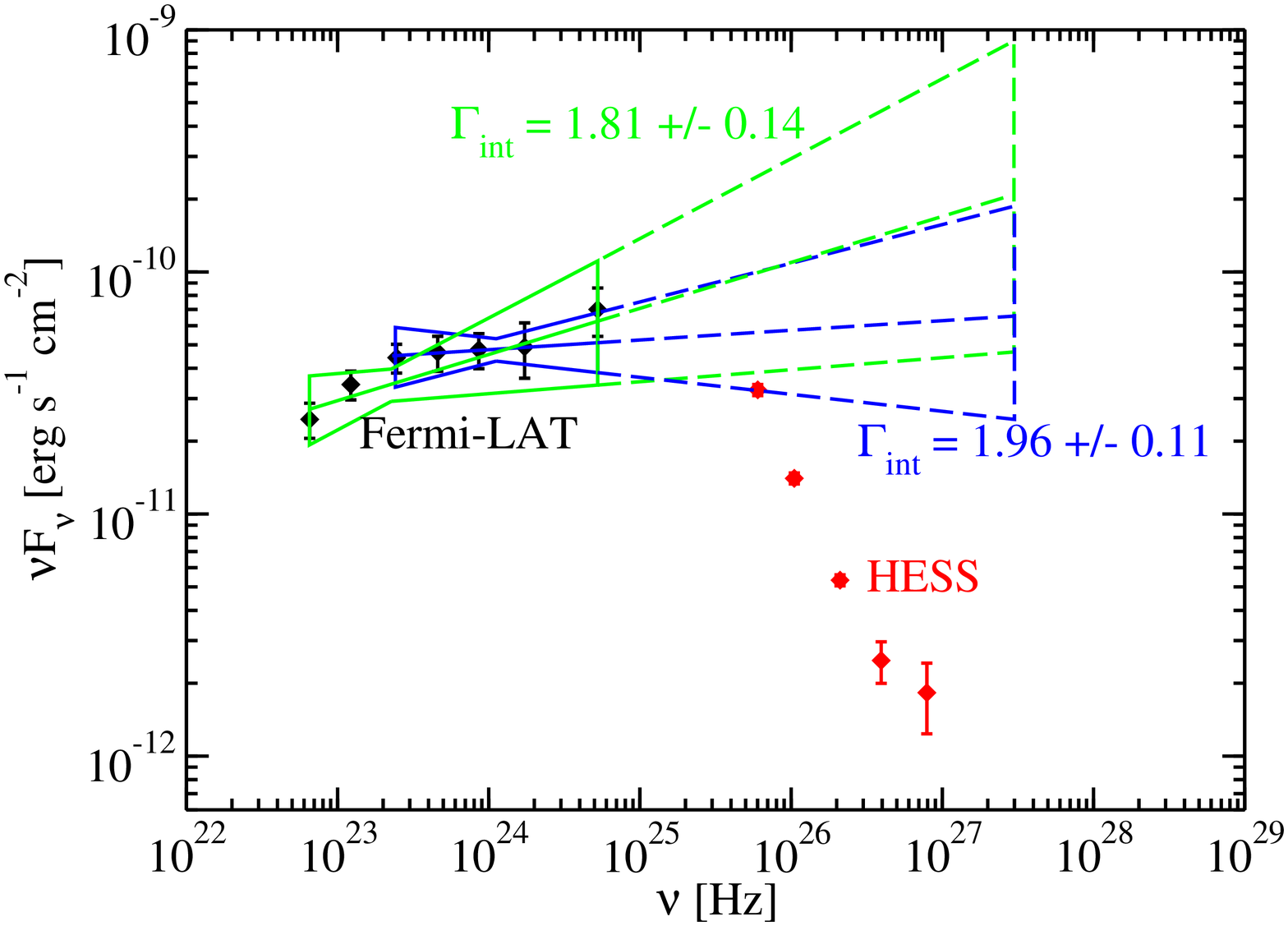}
\caption { The Fermi data (black diamonds) and the extrapolated {\em
Fermi} SED for the two cases described in the text, together with the
{\em HESS} data (red diamonds) for PKS 2155-304. The solid lines
depict the energy range actually covered by LAT, while the broken
lines depict the extrapolation of the LAT SED in the TeV regime.}
\label{SED2155}
\end{figure}

\begin{figure}
\epsscale{0.9}
\includegraphics[scale=0.6]{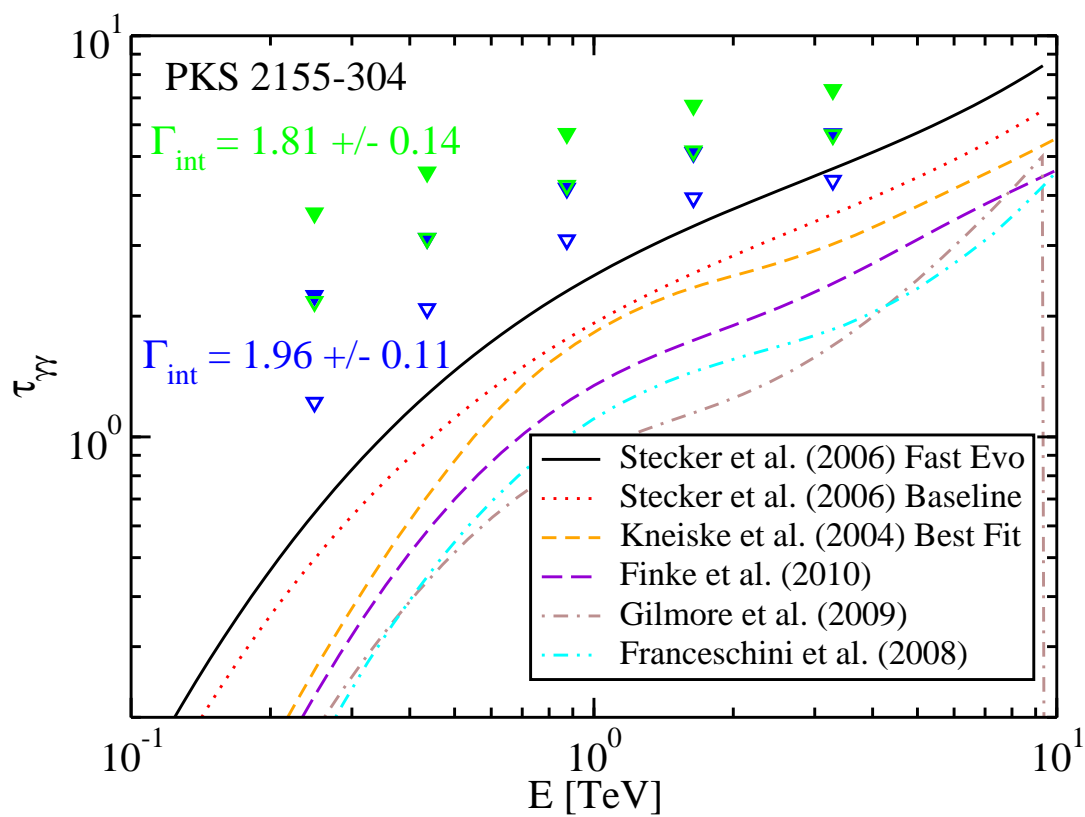}
\caption {The pair production optical depth along the line of sight to
PKS 2155-304 for a range of EBL models, and the constraints our method
imposes on the actual optical depth for the two cases for the GeV
slope described in the text. Empty and solid triangles correspond to
$1\sigma$ and $3\sigma$ upper limits on $\tau_{\gamma\gamma}$.}
\label{tauEBL2155}
\end{figure}

 
 { Methods that constrain the EBL through purely spectral arguments
    are free of the uncertainties of adopting a particular physical model.
    Such methods, based solely on TeV data, have been proposed by 
   \cite{dwek05}, who considered unnatural TeV SEDs that exhibit an exponential increase at their high energy end    and by \cite{schroedter05} who assumed  that all TeV blazars in flaring states have TeV spectra with the same   maximum intrinsic photon index $\Gamma_{TeV}=1.8$.
 Here we present a spectral method for obtaining upper limits to the EBL energy
 density that makes use of simultaneous LAT and TeV observations. 
   In \S \ref{method} we describe our
 method  and apply it to recent simultaneous LAT and TeV observations observations of
 PKS~2155-304,   in \S \ref{disc} we discuss how we can produce  
 stronger constraints on the EBL and demonstrate this using  simultaneous GeV/TeV observations of 1ES 1218+304, and in \S \ref{conc} we conclude.

 \section{An upper limit for the intrinsic TeV SED of TeV blazars
 \label{method}}
 
 Our current observational understanding of blazars points toward a
 SED that consists of two spectral components or more colloquially
 ``bumps''. The first bump peaks at IR to X-ray energies and it is
 almost certainly synchrotron emission from a population of
 relativistic electrons in a partially ordered magnetic field. The
 second bump peaks at MeV to multi-GeV energies and is thought to be
 due to inverse Compton scattering of synchrotron photons
 \citep{band86,bloom96}, a dusty torus \citep{blaz00}, a broad-line
 region \citep{sikora94}, or an accretion disk \citep{dermer92}. For
 the blazars that have been discovered so far, this second bump ranges
 in power between $~0.1-100$ times the power of the synchrotron bump.
  It is also possible
 that the high energy bump is a result of emission from hadrons
 co-accelerated with the jet, which can radiate by hadronic
 synchrotron \citep{muecke01} or photomeson production
 \citep{mannheim92} and the resulting cascades.  Protons may also
 convert to neutrons, escape the relativistic jet, then convert back
 to protons and radiate \citep{atoyan03}.  So far, no 
 observational evidence for a third, higher energy bump has been
 found and  we consider it unlikely that a third SED
 component at TeV energies will be more powerful than the
 extrapolation of the GeV SED at TeV energies.  
   
 Our working assumption is motivated by these considerations and is
 weaker than the statement that there is no third SED bump: {\em we
 assume that the extrapolation of the Fermi-detected SED at TeV
 energies is higher than the intrinsic TeV SED of the source}.  This
 means that the ratio of the actually observed TeV flux $f_{\epsilon,
 obs}$ to the extrapolated one $f_{\epsilon, ext}$ at any given TeV
 energy $\epsilon$, provides a strict upper limit $\tau_{\epsilon,
 max}$ to the EBL-induced pair absorption optical depth at this energy,
\begin{equation}
\label{taumax}
 \tau_{\epsilon, max}=\ln(f_{\epsilon, ext}/f_{\epsilon, obs})\ .
\end{equation} 
 This optical depth can then be compared to the optical depth
 calculated for the various EBL models.  The models for which
 the optical depth is greater that $\tau_{\epsilon, max}$ are then
 excluded.

 An important consideration for our method stems from the mismatch of the
 relatively long time required to obtain a sufficiently good SED of
 typical TeV blazars with {\em Fermi} (no less than $\sim$ a day) to the
 significantly smaller variability time observed by TeV telescopes \citep[in some cases lasting  less than an hour, e.g.][]{aharonian07}. It is clear that, because we want to compare
 simultaneous spectra, the TeV fluxes observed during the rapid TeV
 variability events seen in TeV sources  should not be used together with the {\em Fermi} SED derived after
 integrating for days. 

 \subsection{Application to PKS 2155-304 \label{2155}}
 
 The blazar PKS~2155-304, a high peak frequency BL Lac object at
redshift $z=0.116$, was the target of a multiwavelength campaign in
late August and early September 2008, which included observations by
the {\em Fermi} Gamma-Ray Space Telescope and the HESS atmospheric
Cherenkov telescope \citep{aharonian09}.  It is these observations to
which we now turn our attention.  The $\gamma$-ray SED from this campaign
can be seen in Fig.\ \ref{SED2155}.  \citet{aharonian09} found that
the {\em Fermi} SED derived from data between MJD 54704-54715, the
period of the TeV observations, can be described by a simple power law
with spectral index $\Gamma=1.81\pm0.14$. A longer train of {\em Fermi}
observations between MJD 54682-54743 exhibits a similar GeV state. If
one includes these data, the exposure increases by a factor of $\sim
3.6$ and the preferred GeV spectrum is now a broken power law, whose
high energy part has a photon index $\Gamma=1.96\pm0.11$ \citep{aharonian09}. While a
conservative estimate of the photon index used for obtaining an upper
limit to the TeV flux is that derived from the simultaneous
observations only, the lack of strong variability in the GeV (as well
as the TeV) regime, suggests that it is reasonable to adopt the high
energy photon index as a better description.

The single power-law, with spectral index $\Gamma=1.81\pm0.14$ and high
energy power-law fit, $\Gamma=1.96\pm0.11$ of the broken power-law are
shown as ``bow tie'' error plots in Fig.\ \ref{SED2155}, extrapolated
to the HESS energy range.  These extrapolations are used as upper
limits to the intrinsic flux of the TeV SED, unabsorbed by the EBL.
The upper limit on $\tau_{\gamma\gamma}$ is calculated from equation
(\ref{taumax}) for these two extrapolations, and the results are shown
in Fig.\ \ref{tauEBL2155} along with the absorption optical depth
predictions of several EBL models.  As expected, the steeper GeV index
provides stronger constraints to the EBL models.  Note that the fast
evolution model of Stecker et al. 2006 lies below the $1\sigma$ limit
on $\tau_{\gamma\gamma}$ at the highest TeV energy for $\Gamma=1.96\pm0.11$ and,
in fact, is inconsistent with it at the $1.4\sigma$ level.

\section{How to produce stronger constraints on the EBL  \label{disc}} 

We now discuss how the yield of our method  can be maximized. We base our discussion on the fact that {\em Fermi} is continuously monitoring the
 entire sky, including the extragalactic TeV sources.
 To facilitate our discussion consider  a  source  that has  a LAT spectral index $\alpha_{LAT}$ and a TeV spectral index $\alpha_{TeV}$ (spectral indexes $\alpha$ are connected to photon indexes $\Gamma$ through the relation $\alpha=\Gamma-1$). Let  us further assume that the transition between the two power laws takes place at  an energy $\epsilon_1\approx 100$ GeV, practically the energy border between  LAT and ground based TeV telescopes. Then at an energy $\epsilon_2$ within the TeV regime, the flux one would expect by extrapolating the LAT spectrum is $f_{\epsilon_2, ext}=f_{\epsilon_1} (\epsilon_2/\epsilon_1)^{-\alpha_{LAT}}$, while  the flux $f_{\epsilon_2, obs}$ we actually observe at energy $\epsilon_2$ is  $f_{\epsilon_2, obs}=f_{\epsilon_1} (\epsilon_2/\epsilon_1)^{-\alpha_{TeV}}$, where $f_{\epsilon_1} $ is the flux at energy $\epsilon_1$. This  means that the maximum pair production optical depth $\tau_{\epsilon_2, max}$  that we can infer from observations of this source is
 \begin{equation}
 \tau_{\epsilon_2, max}=(\alpha_{TeV}- \alpha_{LAT}) \ln{\left(\epsilon_2\over \epsilon_1\right)}.
  \end{equation}
Note that since $\tau_{\epsilon_2, max}$ increases with $\epsilon_2$, the higher the TeV energy in which a source is  detected, the stronger the constraints will be.
As expected, the smaller the spectral break $\Delta \alpha=\alpha_{TeV}- \alpha_{LAT}$ of a source is, the stronger the resulting constraint on $\tau_{\epsilon_2, max}$  will be. Therefore, at a given redshift $z$, the most promising sources are those that exhibit the smallest $\Delta\alpha$. If a source varies, states with the smallest  $\Delta\alpha$  will provide the strongest constraints. We discuss this further below.

\subsection{ The benefits of high TeV states}

A general pattern observed in both the low and high energy
bumps of blazar SEDs is that variability events usually manifest
themselves with a hardening of the high energy part/tail of the bump
\citep[e.g.,][]{albert07_mrk421,aharonian09_2155}, with the amplitude
of variability decreasing at progressively lower energies. It is,
therefore, natural to anticipate that when a TeV blazar is in a
high state, the spectrum will harden and rise in amplitude more at TeV
than at GeV energies (variability can explain the fact that in
some variable sources like MRK 421 or S5 0716+714 the extrapolation of
the GeV SED of a given non-flaring epoch to TeV energies
\citep{abdo09b} can be below the actually detected TeV emission at
flaring states).  
This means that in a flaring state the method will provide more severe
limits
on EBL models.

As a potential example of the benefits of catching a source at a
 high state, we return to the case  of PKS 2155-304, whose TeV
flux in the August-September 2008 {\em Fermi}-{\em HESS} campaign was
close to the lowest archival TeV data. 
The source exhibited   a spectral break  $\Delta \alpha=1.38$ ($\alpha_{LAT}=0.96$, $\alpha_{TeV}=2.34$). In a pre-{\sl Fermi} 2006  campaign \citep{aharonian09_2155}, the  TeV spectral index during flaring states hardened up to $\alpha_{TeV}=1.6$. 
 If a similarly hard TeV spectrum is recorded simultaneously with  {\sl Fermi} observations and  if the {\sl Fermi} spectrum does not harden significantly, 
 then $\Delta \alpha=0.74$, which would reduce  $\tau_{max}$ by a factor of 
 $\approx 2$, posing important constraints on the EBL (all the $\tau_{\gamma\gamma}$ upper  limits in Figure \ref{tauEBL2155} would shift downward by a factor of $\sim 2$).

The ideal high states should last long enough to provide us with a
solid determination of the {\em Fermi} flux and photon index. In
addition, the source should remain relatively steady while in the high
flux state, so that the average {\em Fermi} and TeV measurements are a
good representation of the source during this high but not rapidly 
variable state.  A high level of GeV flux is also very useful, because
if there is curvature in the {\em Fermi} band, it is possible that a
broken power law fit will be better than a simple power law and one
can then use the higher energy steeper part of the {\em Fermi} SED to
extrapolate to TeV energies and produce lower values of $\tau_{max}$ 
(as we did for PKS 2155-304).

\subsection{Sources at different redshifts, the case of 1ES 1218+304}

Even if there are sources described by  pure power laws in the entire GeV - TeV regime without the need for a  break,  their observed spectra would be imprinted with a break solely attributed to EBL absorption. Such breaks would increase with increasing redshift. 
 In the more pragmatic case that there is a distribution of intrinsic breaks, then, as the blazar sequence (Fossati et al. 1998) suggests,  the breaks should become stronger  for the more powerful sources seen at higher redshifts, because the peak frequency of the high energy component shifts to lower energies as the source power increases. 
 The increase of $\Delta \alpha$ with
redshift would thus be a convolution of two effects: the
blazar-sequence-like shift of the TeV spectrum to steeper values
(which is intrinsic) and the increase of EBL absorption with
distance.

 If $\Delta \alpha$ is dominated by the former,
our method would  derive its strongest EBL constraints from nearby sources. If, however, the intrinsic increase of the break with redshift is small or negligible, sources at higher redshifts, even with steep TeV spectra  (the most distant example is  3C 279 discovered by  MAGIC \citep{albert08} at $z=0.538$ with $\Gamma_{TeV}=4.11\pm0.68$), would provide us with strong constraints on the EBL.
Because we do not  actually know the intrinsic breaks of the TeV sources,
the question that naturally arises is  under what conditions higher redshift sources with steep TeV spectra
can provide as  useful constraints on the EBL as their nearby siblings that have significantly  harder TeV spectra. 

The important thing to notice here is that  the relevant quantity that constrains the
EBL is  not $\tau_{max}$, but  $\tau_{max} /d_L$, where $d_L$ is the  luminosity distance of the source:
sources with the same value of $\tau_{max} /d_L$ will provide the same constraints on the EBL (assuming that the evolution in the EBL energy density itself with redshift is negligible for small redshifts). This means that, for fixed values of $\epsilon_1$ and $\epsilon_2$, a particular constraint on the EBL is satisfied for sources with a fixed value of $\Delta \alpha / d_L$.
More distant sources with larger breaks $\Delta \alpha$ will provide the same constraints as more nearby sources, as long as they have the same   $\Delta \alpha / d_L$.  As an example, let us use as a reference the state of PKS 2155-304 we discussed above, with $\Delta \alpha=1.38$ and $d_L=533.4$ Mpc.  A source at the distance  of Mrk 421 ($z=0.034$, $d_L=143$ Mpc) will provide the same constraints on the EBL if it exhibits $\Delta \alpha=1.38 \times 143/533.4=0.37$. Any break greater than  $\Delta \alpha=0.37$ will not be as constraining as the stronger  break of the more distant PKS 2155-304.

Going to higher redshifts,  the more distant TeV BL Lac 1ES 1218+304 \citep{albert06}  ($z=0.182$, $d_L=873.5$ Mpc)  would provide the same constraints on the EBL  if it exhibits $\Delta \alpha=1.38\times 873.5/533.4=2.26$. Breaks gentler  than that will produce stronger constraints on the EBL than those derived from the state of PKS 2155-304 we studied here.
 Recently, it has been reported that PKS 1218+304 has been observed  by VERITAS from 2008 December 29 to 2009 April 23 (Acciari et al. 2010).  This is quasi-simultaneous to the LAT observation of the source in the 11 month {\em Fermi}-LAT catalog (A. Abdo et al. 2010 in preperation)\footnote{The catalog is published online at \url{http://fermi.gsfc.nasa.gov/ssc/data/access/lat/1yr\_catalog/} }, which continuously  exposed PKS 1218+304 from 2008 August 4 to 2009 July 4.  These LAT and VERITAS spectra give $\Delta \alpha=1.37$ indicating they can provide a stronger constraint on EBL models.

 Following the same procedure as in the case of PKS 2155-304, we plot in Figure \ref{SED1218} the $\gamma$-ray SED, together with the `bow tie'
extrapolation of the GeV spectrum to VERITAS energies.
 As before, the extrapolation is used as an
upper limit to the intrinsic flux of the TeV SED and the upper limit
on $\tau_{\gamma\gamma}$ is calculated from equation (\ref{taumax})
and plotted in  Fig. \ref{tauEBL1218} along with the absorption
optical depth predictions of several EBL models. 
  As can be seen, {\sl the fast
evolution model of \citet{stecker06} lies  below the $1\sigma$  upper limit on $\tau_{\gamma\gamma}$ for all but the two lowest TeV energies, and below the $3\sigma$ level
 at the highest TeV energy. In particular,  the highest point ($\sim 1.8$ TeV) is inconsistent with Stecker's fast evolution model at $4.7 \sigma$.
  Also the
\citet{stecker06} baseline model, and the \citet{kneiske04} best fit
model are inconsistent at the $2.6\sigma$  level  and $2.9\sigma$ level with $\tau_{max}$  at the highest TeV energy. }}

\section{Conclusions \label{conc}}

We presented a simple, model-independent method for setting upper limits to the EBL. Our method is based on the assumption that the level of the intrinsic TeV emission of blazars is below the extrapolation of the LAT SED to TeV energies. We applied our method to PKS 2155-304 and 1ES 1218+304, the only two TeV blazars of known redshift for which  published  simultaneous LAT-TEV observations currently exist. Even with these first  applications of our method, the highest level EBL models are disfavored. Future LAT-TeV simultaneous  observations hold the promise of pushing the EBL upper limits much further.
We argued that it is important to devote TeV time not only to nearby sources, but also to more distant sources, hoping in both cases to observe high TeV states that for a given source will exhibit  smaller GeV to TeV spectral break $\Delta \alpha$, and will, therefore, produce stronger constraints. 

 Because the most difficult observations to obtain are in the TeV
band, requests to monitor particular sources with TeV facilities can
be triggered from {\sl Fermi} observations of high states of TeV
sources.  A complementary approach is X-ray monitoring of the TeV
blazars.  In this case, because the X-ray emission of most TeV blazars
is the high energy tail of the synchrotron component,
  high X-ray states are, in general,  a good proxy
for  high TeV states (e.g. \cite{aharonian09_2155}).  Such X-ray
monitoring holds the promise of catching states in which, while the
GeV emission does not increase substantially, the TeV emission does.

We note here that our method assumes that the entire spectral break from the LAT to the TeV bands is due to EBL  pair production absorption. This is an extreme assumption and it is highly probable that a substantial fraction of the break is intrinsic to the source. 
This in turn means that the actual level of the EBL may be significantly lower than the upper limits produced by our method. 
It would be very exciting and possibly hinting to new physics
\citep[e.g.,][]{amelino01} if the lowest collective values of
 $\tau_{max}$ that our method will produce,
challenge the lower level on the EBL inferred by galaxy counts
\citep[e.g.,][]{madau00, fazio04,bethermin10}. We anticipate that current and upcoming TeV-GeV blazar monitoring campaigns will provide plenty of opportunity for applying our method.

\acknowledgements

We thank  Luigi Costamante for useful discussions.
MG acknowledges support from the NASA grants ATFP NNX08AG77G and {\sl Fermi} NNX09AR88G. JDF was
partially supported by NASA Swift Guest Investigator Grant
DPR-NNG05ED411 and NASA GLAST Science Investigation DPR-S-1563-Y.
LCR acknowledges the support by the Kavli Institute for Cosmological
Physics at the University of Chicago through grants NSF PHY-
0114422 and NSF PHY-0551142 and an endowment from the Kavli
Foundation and its founder Fred Kavli.

\clearpage

\begin{figure}
\epsscale{0.9}
\includegraphics[scale=0.6]{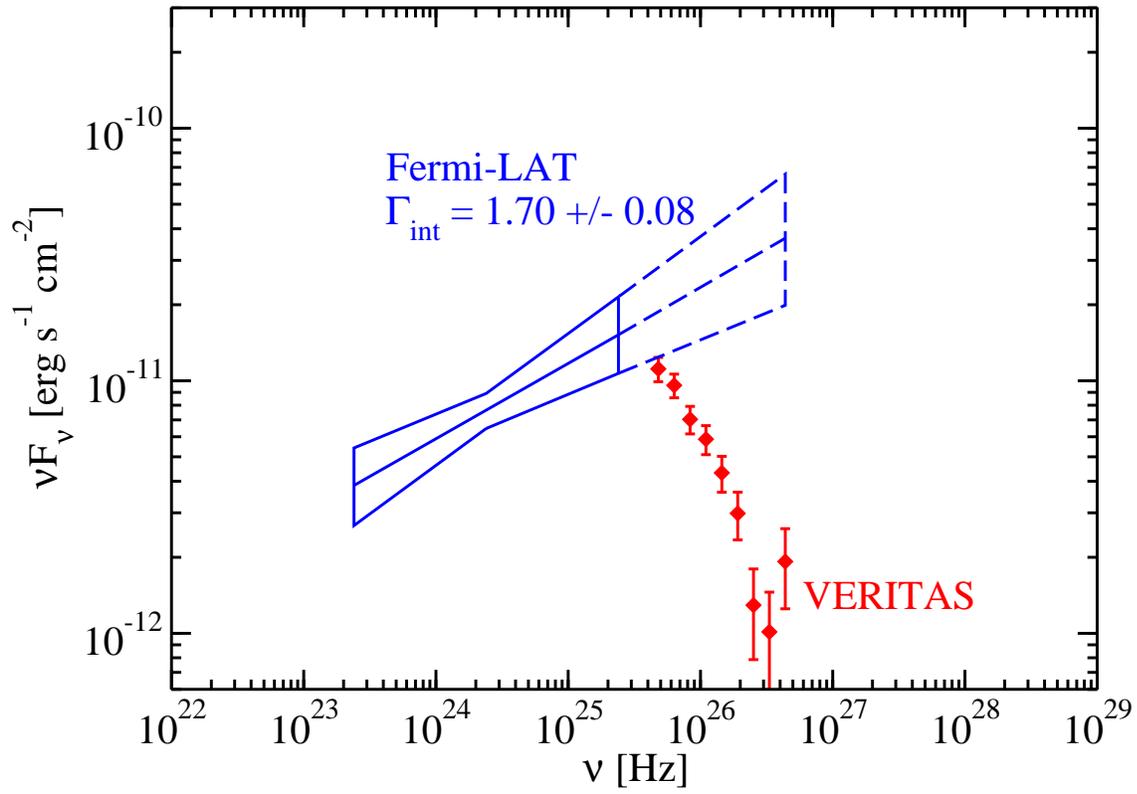}
\caption { The {\sl Fermi}-LAT SED ({\sl Fermi} 1 year on-line LAT
catalog; solid lines) extrapolated to TeV energies (broken lines),
together with the VERITAS data (red diamonds) for 1ES 1218+304.}
\label{SED1218}
\end{figure}

\begin{figure}
\epsscale{0.9}
\includegraphics[scale=0.6]{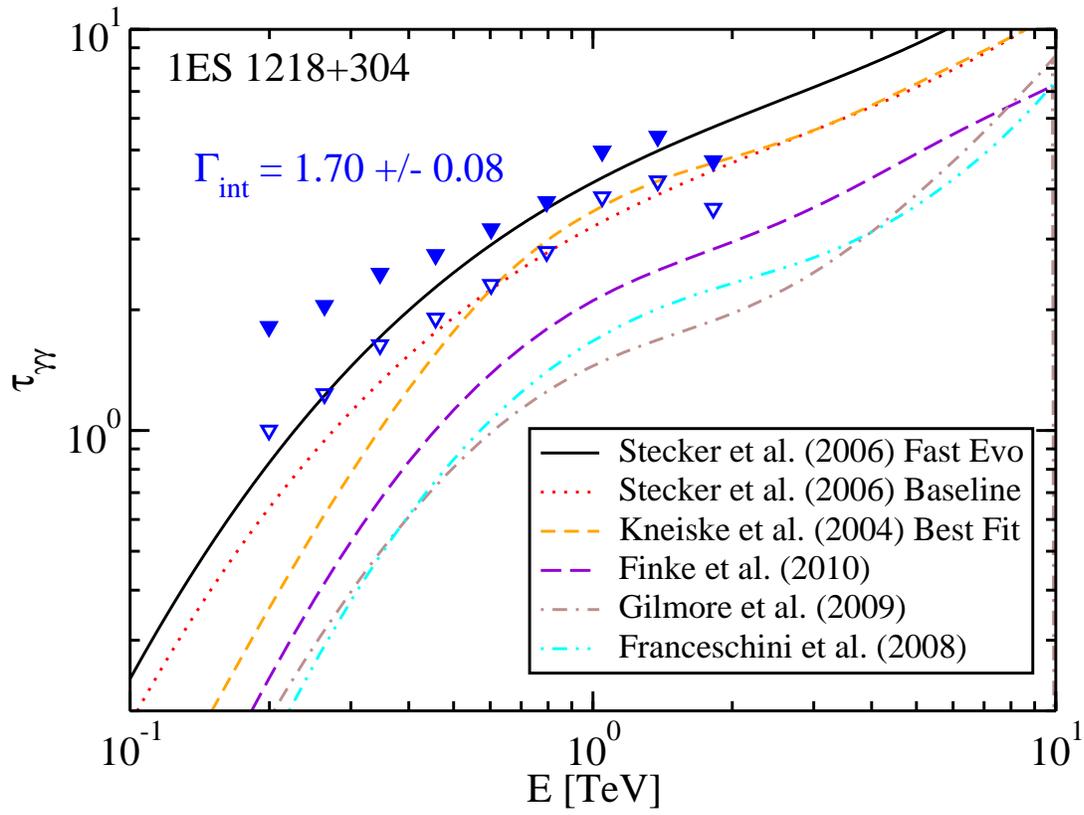}
\caption {Same as in Fig. \ref{tauEBL2155}, but for 1ES 1218+304.}
\label{tauEBL1218}
\end{figure}

\end{document}